%% file: pro2.tex
\documentclass[FBSedit,FBSmath,ecsub]{FBSsuppl}
\usepackage{amsfonts}
\usepackage{amssymb}

\title{A Practical Method for  
Relativistic 3N-Scattering Calculations with Realistic Potentials}
\author{H. Kamada\thanks{\textit{Alternative address:} Institut f\"ur Strahlen und Kernphysik der 
Universit\"at Bonn, D53115 Bonn, Germany
}\comma\thanks{\textit{E-mail address:} 
kamada@hadron.tp2.ruhr-uni-bochum.de}}           
\institute{Institut f\"ur
 theoretische Physik II, Ruhr-Universit\"at Bochum, \\
D44780 Bochum, Germany}

\begin{document}

\maketitle
\begin{abstract}
Using a scale transformation in momentum space a phase equivalent 
relativistic potential is generated from the nonrelativistic 
potential. By that transformation 
 a practical method for the relativistic 3N scattering with 
realistic nucleon-nucleon potentials is introduced. 
The formalism
can be  applied to any realistic nonrelativistic potential.
We also discuss the locations of the moving logarithmic singularities of 
the free relativistic
three-body Green's function.
It enters in the relativistic 3N Faddeev equations, which have been formulated 
long time ago for a 3-boson bound state and which we propose to also use for 
3N scattering. Finally we compare relativistic deuteron wave functions to 
nonrelativistic ones. 
\end{abstract}

\section{Introduction}
 
Recently the total cross section for nd scattering has been measured 
for 
incident neutron energies from 50 MeV to 600MeV  \cite{Abfalterer98}.
The data have been analyzed\cite{Witala99}
up to 300 MeV by rigorous 3N Faddeev calculations\cite{GloeckleReport} based on 
the CD-Bonn 
potential\cite{Machleidt}  and the Tucson-Melbourne three-nucleon force (3NF)\cite{TM}. Two nucleon forces alone are not sufficient to describe the data 
above about 100 MeV.
The discrepancy is very likely filled by 3NF effects and relativistic corrections. 
In \cite{Witala99} we simply estimated one relativistic kinematical effect related to the incident flux. 
This leads to an increase of the total cross section
by  about 3\% at 100MeV and about 7\% at 250MeV.
The rest of the discrepancy of about the  same  amount turned out to be 
understandable as a 3NF contribution. 3NF effects can be seen already at 
lower energies around 65 MeV nucleon lab energy in the minimum of the differential
cross section in elastic pd scattering. This is more pronounced at higher 
energies as recent measurements at RIKEN\cite{Sakamoto,Sekiguchi} and  IUCF\cite{Rohdjess} show. 
Those discrepancies to NN force predictions alone are known as Sagara discrepancy\cite{Sagara,Nemoto,Bochum}.
This very simplistic estimate of a relativistic correction just mentioned is not 
a substitute for a consistent treatment. The dynamical equation should be relativistic, not only the kinematics. 

 Relativistic effects in a three-boson bound state
have been  discussed in\cite{Gloeckle86}.
Following those concepts several questions arise: \\
(1) How do we formulate the relativistic three-body equations for scattering 
processes? \\ 
(2) To what extent does relativity  
require modifications of the conventional nuclear forces? 

With respect to  (1) the answer is given in\cite{Gloeckle86} for  three bosons 
by appropriately supplementing the relativistic Faddeev equation presented in \cite{Gloeckle86} for scattering. 
An important ingredient in these equations is 
the boost transformation of the two-body t-matrices. What is not treated in \cite{Gloeckle86} 
are the Wigner rotations of the nucleon spin. This has to be added.
This formalism in \cite{Gloeckle86} is of the instant form of relativistic 
dynamics and has been proposed  by F. Coester in\cite{Coester65}.

 In order to answer the second question one has to introduce new forces 
which inserted into the relativistic two-body Schr\"odinger equation reproduce the
two-body observables 
as precisely as the  nonrelativistic potentials.

Recently Gl\"ockle and the author
 found an analytical momentum-transformation\cite{KGT} which provides 
 a mathematical relation between the 
relativistic and nonrelativistic two-body Schr\"odinger equation and thus between 
the potentials in the two equations. 
This transformation is such that the NN phase shifts and the binding energy of the
deuteron do not change. 
In \cite{Urbana} the parameters of a given NN potential (AV18) have been readjusted when used together with the relativistic form of the kinetic energy in order to guarantee the same NN phase shifts to some degree of accuracy.
Our transformation given in \cite{KGT} guarantees exactly the same phase shifts and no refitting is required.
These tranformed NN forces will be the dynamical input for the properly 
supplemented relativistic Faddeev equation in our treatment of 
relativistic 3N scattering.

In section 2 we will briefly review the formalism proposed in \cite{Gloeckle86}
for a relativistic 3N Faddeev equation and extend it from bound state to scattering 
calculations. 
Thereby the treatment of the logarithmic moving 
singularities arising from the free 3N propagator requires special considerations. 
In section 3 we address the transformation which generates from the nonrelativistic 
NN potential a new one adequate to the relativistic form 
of the Schr\"odinger equation and compare the resulting relativistic deuteron 
wave function with the one established in a recent study of the Urbana group
\cite{Urbana}.
The summary is given in section 4.

\section{ Gl\"ockle - Lee - Coester Relativistic Faddeev Equation}

In the article \cite{Gloeckle86} a relativistic Faddeev equation is introduced 
for a  bound state of three bosons. 
In the Bakamjian-Thomas scheme \cite{BakaThomas} the two-body potential $v_{ij}$
which is defined in the two-body c.m. system (2CM), is transformed into 
$V_{ij}$ which belongs to the 
three-body c.m. system (3CM), as 
\begin{eqnarray}
V_{ij}\equiv \sqrt{ (\omega _{ij} +v_{ij} )^ 2 + \vec p _{ij} ^2 }
- \sqrt{ \omega_{ij} ^ 2 + \vec p_{ij} ^ 2 }  
\label{e1}
\end{eqnarray}
where $\omega_{ij} $$=2 \sqrt{m^2 +\vec k^ 2 }$ and $\vec p_{ij}$ are the 
energy and total 
momentum of the noninteractiong pair of particles $ij$. 
The usual nonrelativistic relative momentum $ {1 \over 2 } 
( \vec k_i -\vec k_j)$, where $\vec k_{i,j}$ are the individual momenta, 
is replaced now by the 
Lorentz tranformed individual momenta $\vec k $ and $-\vec k$ in the 2CM :
\begin{eqnarray}
\vec k = { 1 \over 2 } (\vec k_i - \vec k_j )  
-{\vec p_{ij} \over 2 } { { E_i - E_j } \over { E_i + E_j +
 \sqrt{ (E_i + E_j )^2 - \vec p _{ij} ^2 } }}
\end{eqnarray}
The individual energies are 
$E_i= \sqrt{ m^ 2 + k_i^2 }$.
The "boosted" potential $V_{ij} $ depends on the total momentum 
of the two-body subsystem as is obvious from Eq. (\ref{e1}). 
Using the complete set of two-body states,  $ < \vec k \vert \phi_b  >$ and
$< \vec k \vert \phi (k_0) > \equiv <\vec k \vert \vec k_0 > ^ {(+)} $,
$V_{ij}$ is explicitly given as , 
\begin{eqnarray}
&&~~~~~~~~~~~~~~~~~~~~~~~~~~~~< \vec p_{ij} \vec k \vert V_{ij} \vert {\vec p_{ij}}' {\vec k}' > =
\cr &&
\delta (\vec p_{ij} -{ \vec p_{ij} }' ) \int d{\vec k_0}  < \vec k \vert \vec k_0 > ^ {(+)} 
\sqrt{ \omega^2 (k_0) + \vec p_{ij} ^ 2 } ~~^{(+)}<\vec k_0 \vert {\vec k }' > 
\cr && 
-\delta (\vec p_{ij} -{ \vec p_{ij} }' )
 \delta (\vec k -{ \vec k }' ) \sqrt{\omega ^2 ( k )+ \vec p_{ij} ^2 }  +  \dots 
\label{e2}
\end{eqnarray} 
where 
\begin{eqnarray}
<\vec k \vert ( \omega (k) + v _{ij} ) \vert \phi (\vec k_0 )> 
 = \omega (k_0) < \vec k \vert \phi (k_0)>.
\label{e3}
\end{eqnarray}
The dots denote the two-body bound state contribution. 
Instead of using (\ref{e2}) one can also express directly 
 the two-body boosted transition operator $T_{ij}$ in the 3CM by  
the t-matrix in the 2CM, as shown in\cite{Gloeckle86}. 

The free three-body Green´s function is singular above the 
three-body breakup threshold. It can be written as 
\begin{eqnarray}
G_0^ {-1} &=&  E- \sqrt{ \omega ^2 (k) + \vec p_{ij}^2 } - \sqrt{\vec p_{ij}^2 + m^2 } + i \epsilon
 \cr  &=&  E - \sqrt{ \vec p _{ij} ^2 + m^ 2 } - \sqrt{ \vec p_{jk} ^2 + m^ 2} 
- \sqrt  {\vec  p_{ij} ^2 + \vec p_{jk} ^2 + 2 p_{ij} p_{jk} x + m^ 2 } 
\end{eqnarray} 
which exhibits an angular dependence ($ x = \hat p_{ij} \cdot  \hat p_{kj} $) 
After integration over  $x$ logarithmic singularities appear in the Faddeev equation.
By substituting  $ y=G_0 ^ {-1} (x) $ the integration can be carried through 
with the result $ \int dx/y  = [ y +
 ( \sqrt{\vec p_{ij} ^2 + m^2 }
+\sqrt{\vec p_{jk}^2  + m^2 } - E  ~) ln (y)  ] /p_{ij}p_{jk}$. 
The singularities occur under the conditions 
$x=\pm1$ and $G_0^{ -1}=0$. Their locations are displayed
 in Fig. 1 for the example of $E_{lab}$ = 250 MeV. We see a shift of the relativistic singularity lines
in comparison to the nonrelativistic ones. 

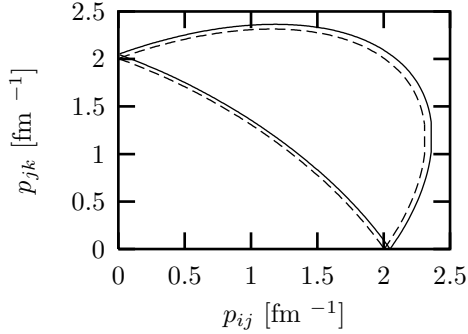
\begin{figure}[hbt]
\input{fig1.tex}
\caption[]{ The loci of the Green's function singularity at $E_{lab}=$250 MeV.
The solid (dashed) line is for relativistic (nonrelativistic) kinematics.
}
\end{figure}

\section{  Momentum Scale Transformation }

In \cite{KGT} a momentum scale transformation has been introduced by  which the nonrelativistic potential $v^ {NR}$ can be rewritten into a "relativistic" one
, $v$, as used in Eq. (\ref{e1}) without changing the on-the-energy-shell properties in the two-body 
system. The nonrelativistic potential $v^{NR}$ enters the nonrelativistic 
Schr\"odinger equation,
\begin{eqnarray}
<\vec k ^{NR} \vert ( {{ \vec k ^ {NR} }^2 \over m} + 2m + v^{NR}_{ij} ) \vert \psi (k_0 ^ {NR} ) >
= ( {{ k_0^{NR}} ^2 \over m } + 2 m )  < { \vec k^ {NR} }  \vert \psi (k_0 ^ {NR})  > 
\end{eqnarray}
The corresponding relativistic equation is given in Eq. (\ref{e3}). 
The momentum transformation connecting relativistic and nonrelativistic momenta.
\begin{eqnarray}
\omega (k) = 2 \sqrt{ k ^2 + m^2 } \equiv  {{ k^{NR}} ^2 \over m } + 2 m
\label{e4}
\end{eqnarray}
leads to the relation between the relativistic and nonrelativistic wave functions:
\begin{eqnarray} 
\phi( k ) =  { \sqrt{2 } m \over \sqrt{ \sqrt{ m^2 + k ^2 } \sqrt{ 2m( m +  \sqrt{m^2 + k^2} )
 } }}  \psi ^ {NR} ( \sqrt {2m ( \sqrt { m^2 + k ^2 }- m)  }  ).
\label{e5}
\end{eqnarray}
Fig. 2. shows the S- and D- wave components of the deuteron wave function
for the AV18 \cite{AV18} model potential in configuration space. 

At short distances the relativistic  wave function becomes larger, 
a  tendency which agrees with the result  of a recent study 
\cite{Urbana}. In \cite{Urbana} an approximation to the ratio between the relativistic 
and  
nonrelativistic  D- wave functions ( see Eq. (2.27) )has been considered: 
\begin{eqnarray}
R \equiv { \phi _D ( k ) \over \psi^ {NR} _D ( k ) } = 
{  k^2 \over { 2 (\sqrt { m^ 2 + k ^2 } -m ) \sqrt{m^2 + k^2 }      }} 
\label{e6}
\end{eqnarray}
We compare that ratio to the one given in Eq.(\ref{e5}) in Fig. 3 
and thereby neglect the momentum transformation of the argument in $\psi ^ {NR}$. 
Both results agree well, especially if one takes into account, that Eq. (\ref{e6}) 
is only an approximation  and that our result is even closer to the real calculated
ratio in \cite{Urbana}. 
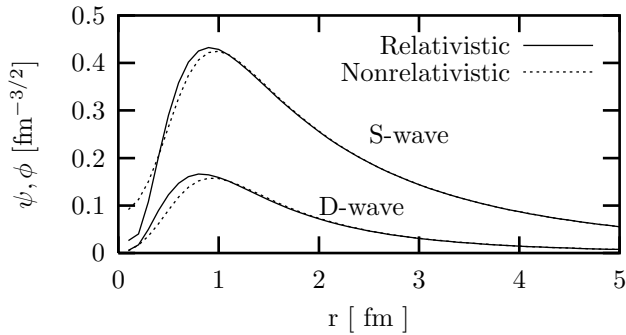
\begin{figure}[hbt]
\input{fig2.tex}
\caption[]{ The deuteron wave functions.  The solid (dotted) lines are 
the relativistic 
(nonrelativistic) wave functions.   }
\end{figure}
\begin{figure}[hbt]
\input{fig3.tex}
\caption[]{ The ratio between the relativistic D- wave deuteron wave function and the 
nonrelativistic one. The solid and dashed curves are due to 
Eq. (\ref{e5}) and Eq. (\ref{e6}), respectively. }
\end{figure}
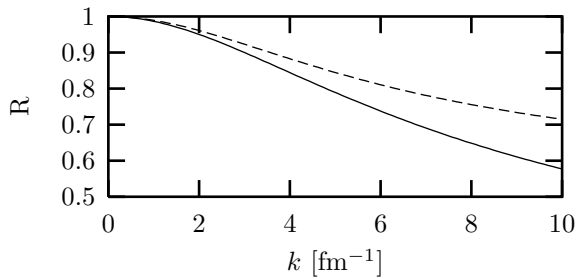 

\section{ Summary }

We reviewed briefly the formalism in\cite{Gloeckle86},which was devoted to a relativistic 
treatment of a three-boson bound state in the instant form of relativistic dynamics. 
The relativistic Faddeev equation  given there is extended to three-nucleon 
scattering and the necessary changes in the free propagator singularities are 
pointed out.
This scheme requires NN forces which are tuned to NN phaseshifts together 
with the relativistic form of the kinetic energy.
We proposed to use the momentum scale transformation from\cite{KGT}, which 
provides an analytical connection between the potentials in the nonrelativistic 
and relativistic two-body Schr\"odinger equations. 
No refitting like in \cite{Urbana} is required. 
We did not yet address the Wigner rotation of the spin states, but we expect 
this to be achieved along the line given in \cite{Polyzou}.
We also plan to proceed without partial wave decomposition and then the first 
steps formulated by \cite{Fachruddin} will be useful.

\begin{acknowledge}
I would like to dedicate this paper to Prof. Walter Gl\"ockle on the occasion of his 60th 
birthday.  
I would like to thank Prof. Hideyuki Sakai and  Prof. Kichiji Hatanaka 
for fruitful 
discussions of the 3N scattering measurements at  RIKEN and RCNP.
It was a pleasure for me to join  
Prof. Shinsho Oryu, Prof. Masayasu Kamimura, 
Prof. Souichi Ishikawa and  Prof. Yasuro Koike 
 in the organization of 
the first Asia-Pacific Few-Body conference.
This work is financially supported by the 
Deutsche Forschungsgemeinschaft. 
\end{acknowledge}

\end{document}

%% file: fig1.tex
\begingroup%
  \makeatletter%
  \newcommand{\GNUPLOTspecial}{%
    \@sanitize\catcode`\%=14\relax\special}%
  \setlength{\unitlength}{0.1bp}%
{\GNUPLOTspecial{!
/gnudict 256 dict def
gnudict begin
/Color false def
/Solid false def
/gnulinewidth 5.000 def
/userlinewidth gnulinewidth def
/vshift -33 def
/dl {10 mul} def
/hpt_ 31.5 def
/vpt_ 31.5 def
/hpt hpt_ def
/vpt vpt_ def
/M {moveto} bind def
/L {lineto} bind def
/R {rmoveto} bind def
/V {rlineto} bind def
/vpt2 vpt 2 mul def
/hpt2 hpt 2 mul def
/Lshow { currentpoint stroke M
  0 vshift R show } def
/Rshow { currentpoint stroke M
  dup stringwidth pop neg vshift R show } def
/Cshow { currentpoint stroke M
  dup stringwidth pop -2 div vshift R show } def
/UP { dup vpt_ mul /vpt exch def hpt_ mul /hpt exch def
  /hpt2 hpt 2 mul def /vpt2 vpt 2 mul def } def
/DL { Color {setrgbcolor Solid {pop []} if 0 setdash }
 {pop pop pop Solid {pop []} if 0 setdash} ifelse } def
/BL { stroke gnulinewidth 2 mul setlinewidth } def
/AL { stroke gnulinewidth 2 div setlinewidth } def
/UL { gnulinewidth mul /userlinewidth exch def } def
/PL { stroke userlinewidth setlinewidth } def
/LTb { BL [] 0 0 0 DL } def
/LTa { AL [1 dl 2 dl] 0 setdash 0 0 0 setrgbcolor } def
/LT0 { PL [] 1 0 0 DL } def
/LT1 { PL [4 dl 2 dl] 0 1 0 DL } def
/LT2 { PL [2 dl 3 dl] 0 0 1 DL } def
/LT3 { PL [1 dl 1.5 dl] 1 0 1 DL } def
/LT4 { PL [5 dl 2 dl 1 dl 2 dl] 0 1 1 DL } def
/LT5 { PL [4 dl 3 dl 1 dl 3 dl] 1 1 0 DL } def
/LT6 { PL [2 dl 2 dl 2 dl 4 dl] 0 0 0 DL } def
/LT7 { PL [2 dl 2 dl 2 dl 2 dl 2 dl 4 dl] 1 0.3 0 DL } def
/LT8 { PL [2 dl 2 dl 2 dl 2 dl 2 dl 2 dl 2 dl 4 dl] 0.5 0.5 0.5 DL } def
/Pnt { stroke [] 0 setdash
   gsave 1 setlinecap M 0 0 V stroke grestore } def
/Dia { stroke [] 0 setdash 2 copy vpt add M
  hpt neg vpt neg V hpt vpt neg V
  hpt vpt V hpt neg vpt V closepath stroke
  Pnt } def
/Pls { stroke [] 0 setdash vpt sub M 0 vpt2 V
  currentpoint stroke M
  hpt neg vpt neg R hpt2 0 V stroke
  } def
/Box { stroke [] 0 setdash 2 copy exch hpt sub exch vpt add M
  0 vpt2 neg V hpt2 0 V 0 vpt2 V
  hpt2 neg 0 V closepath stroke
  Pnt } def
/Crs { stroke [] 0 setdash exch hpt sub exch vpt add M
  hpt2 vpt2 neg V currentpoint stroke M
  hpt2 neg 0 R hpt2 vpt2 V stroke } def
/TriU { stroke [] 0 setdash 2 copy vpt 1.12 mul add M
  hpt neg vpt -1.62 mul V
  hpt 2 mul 0 V
  hpt neg vpt 1.62 mul V closepath stroke
  Pnt  } def
/Star { 2 copy Pls Crs } def
/BoxF { stroke [] 0 setdash exch hpt sub exch vpt add M
  0 vpt2 neg V  hpt2 0 V  0 vpt2 V
  hpt2 neg 0 V  closepath fill } def
/TriUF { stroke [] 0 setdash vpt 1.12 mul add M
  hpt neg vpt -1.62 mul V
  hpt 2 mul 0 V
  hpt neg vpt 1.62 mul V closepath fill } def
/TriD { stroke [] 0 setdash 2 copy vpt 1.12 mul sub M
  hpt neg vpt 1.62 mul V
  hpt 2 mul 0 V
  hpt neg vpt -1.62 mul V closepath stroke
  Pnt  } def
/TriDF { stroke [] 0 setdash vpt 1.12 mul sub M
  hpt neg vpt 1.62 mul V
  hpt 2 mul 0 V
  hpt neg vpt -1.62 mul V closepath fill} def
/DiaF { stroke [] 0 setdash vpt add M
  hpt neg vpt neg V hpt vpt neg V
  hpt vpt V hpt neg vpt V closepath fill } def
/Pent { stroke [] 0 setdash 2 copy gsave
  translate 0 hpt M 4 {72 rotate 0 hpt L} repeat
  closepath stroke grestore Pnt } def
/PentF { stroke [] 0 setdash gsave
  translate 0 hpt M 4 {72 rotate 0 hpt L} repeat
  closepath fill grestore } def
/Circle { stroke [] 0 setdash 2 copy
  hpt 0 360 arc stroke Pnt } def
/CircleF { stroke [] 0 setdash hpt 0 360 arc fill } def
/C0 { BL [] 0 setdash 2 copy moveto vpt 90 450  arc } bind def
/C1 { BL [] 0 setdash 2 copy        moveto
       2 copy  vpt 0 90 arc closepath fill
               vpt 0 360 arc closepath } bind def
/C2 { BL [] 0 setdash 2 copy moveto
       2 copy  vpt 90 180 arc closepath fill
               vpt 0 360 arc closepath } bind def
/C3 { BL [] 0 setdash 2 copy moveto
       2 copy  vpt 0 180 arc closepath fill
               vpt 0 360 arc closepath } bind def
/C4 { BL [] 0 setdash 2 copy moveto
       2 copy  vpt 180 270 arc closepath fill
               vpt 0 360 arc closepath } bind def
/C5 { BL [] 0 setdash 2 copy moveto
       2 copy  vpt 0 90 arc
       2 copy moveto
       2 copy  vpt 180 270 arc closepath fill
               vpt 0 360 arc } bind def
/C6 { BL [] 0 setdash 2 copy moveto
      2 copy  vpt 90 270 arc closepath fill
              vpt 0 360 arc closepath } bind def
/C7 { BL [] 0 setdash 2 copy moveto
      2 copy  vpt 0 270 arc closepath fill
              vpt 0 360 arc closepath } bind def
/C8 { BL [] 0 setdash 2 copy moveto
      2 copy vpt 270 360 arc closepath fill
              vpt 0 360 arc closepath } bind def
/C9 { BL [] 0 setdash 2 copy moveto
      2 copy  vpt 270 450 arc closepath fill
              vpt 0 360 arc closepath } bind def
/C10 { BL [] 0 setdash 2 copy 2 copy moveto vpt 270 360 arc closepath fill
       2 copy moveto
       2 copy vpt 90 180 arc closepath fill
               vpt 0 360 arc closepath } bind def
/C11 { BL [] 0 setdash 2 copy moveto
       2 copy  vpt 0 180 arc closepath fill
       2 copy moveto
       2 copy  vpt 270 360 arc closepath fill
               vpt 0 360 arc closepath } bind def
/C12 { BL [] 0 setdash 2 copy moveto
       2 copy  vpt 180 360 arc closepath fill
               vpt 0 360 arc closepath } bind def
/C13 { BL [] 0 setdash  2 copy moveto
       2 copy  vpt 0 90 arc closepath fill
       2 copy moveto
       2 copy  vpt 180 360 arc closepath fill
               vpt 0 360 arc closepath } bind def
/C14 { BL [] 0 setdash 2 copy moveto
       2 copy  vpt 90 360 arc closepath fill
               vpt 0 360 arc } bind def
/C15 { BL [] 0 setdash 2 copy vpt 0 360 arc closepath fill
               vpt 0 360 arc closepath } bind def
/Rec   { newpath 4 2 roll moveto 1 index 0 rlineto 0 exch rlineto
       neg 0 rlineto closepath } bind def
/Square { dup Rec } bind def
/Bsquare { vpt sub exch vpt sub exch vpt2 Square } bind def
/S0 { BL [] 0 setdash 2 copy moveto 0 vpt rlineto BL Bsquare } bind def
/S1 { BL [] 0 setdash 2 copy vpt Square fill Bsquare } bind def
/S2 { BL [] 0 setdash 2 copy exch vpt sub exch vpt Square fill Bsquare } bind def
/S3 { BL [] 0 setdash 2 copy exch vpt sub exch vpt2 vpt Rec fill Bsquare } bind def
/S4 { BL [] 0 setdash 2 copy exch vpt sub exch vpt sub vpt Square fill Bsquare } bind def
/S5 { BL [] 0 setdash 2 copy 2 copy vpt Square fill
       exch vpt sub exch vpt sub vpt Square fill Bsquare } bind def
/S6 { BL [] 0 setdash 2 copy exch vpt sub exch vpt sub vpt vpt2 Rec fill Bsquare } bind def
/S7 { BL [] 0 setdash 2 copy exch vpt sub exch vpt sub vpt vpt2 Rec fill
       2 copy vpt Square fill
       Bsquare } bind def
/S8 { BL [] 0 setdash 2 copy vpt sub vpt Square fill Bsquare } bind def
/S9 { BL [] 0 setdash 2 copy vpt sub vpt vpt2 Rec fill Bsquare } bind def
/S10 { BL [] 0 setdash 2 copy vpt sub vpt Square fill 2 copy exch vpt sub exch vpt Square fill
       Bsquare } bind def
/S11 { BL [] 0 setdash 2 copy vpt sub vpt Square fill 2 copy exch vpt sub exch vpt2 vpt Rec fill
       Bsquare } bind def
/S12 { BL [] 0 setdash 2 copy exch vpt sub exch vpt sub vpt2 vpt Rec fill Bsquare } bind def
/S13 { BL [] 0 setdash 2 copy exch vpt sub exch vpt sub vpt2 vpt Rec fill
       2 copy vpt Square fill Bsquare } bind def
/S14 { BL [] 0 setdash 2 copy exch vpt sub exch vpt sub vpt2 vpt Rec fill
       2 copy exch vpt sub exch vpt Square fill Bsquare } bind def
/S15 { BL [] 0 setdash 2 copy Bsquare fill Bsquare } bind def
/D0 { gsave translate 45 rotate 0 0 S0 stroke grestore } bind def
/D1 { gsave translate 45 rotate 0 0 S1 stroke grestore } bind def
/D2 { gsave translate 45 rotate 0 0 S2 stroke grestore } bind def
/D3 { gsave translate 45 rotate 0 0 S3 stroke grestore } bind def
/D4 { gsave translate 45 rotate 0 0 S4 stroke grestore } bind def
/D5 { gsave translate 45 rotate 0 0 S5 stroke grestore } bind def
/D6 { gsave translate 45 rotate 0 0 S6 stroke grestore } bind def
/D7 { gsave translate 45 rotate 0 0 S7 stroke grestore } bind def
/D8 { gsave translate 45 rotate 0 0 S8 stroke grestore } bind def
/D9 { gsave translate 45 rotate 0 0 S9 stroke grestore } bind def
/D10 { gsave translate 45 rotate 0 0 S10 stroke grestore } bind def
/D11 { gsave translate 45 rotate 0 0 S11 stroke grestore } bind def
/D12 { gsave translate 45 rotate 0 0 S12 stroke grestore } bind def
/D13 { gsave translate 45 rotate 0 0 S13 stroke grestore } bind def
/D14 { gsave translate 45 rotate 0 0 S14 stroke grestore } bind def
/D15 { gsave translate 45 rotate 0 0 S15 stroke grestore } bind def
/DiaE { stroke [] 0 setdash vpt add M
  hpt neg vpt neg V hpt vpt neg V
  hpt vpt V hpt neg vpt V closepath stroke } def
/BoxE { stroke [] 0 setdash exch hpt sub exch vpt add M
  0 vpt2 neg V hpt2 0 V 0 vpt2 V
  hpt2 neg 0 V closepath stroke } def
/TriUE { stroke [] 0 setdash vpt 1.12 mul add M
  hpt neg vpt -1.62 mul V
  hpt 2 mul 0 V
  hpt neg vpt 1.62 mul V closepath stroke } def
/TriDE { stroke [] 0 setdash vpt 1.12 mul sub M
  hpt neg vpt 1.62 mul V
  hpt 2 mul 0 V
  hpt neg vpt -1.62 mul V closepath stroke } def
/PentE { stroke [] 0 setdash gsave
  translate 0 hpt M 4 {72 rotate 0 hpt L} repeat
  closepath stroke grestore } def
/CircE { stroke [] 0 setdash 
  hpt 0 360 arc stroke } def
/Opaque { gsave closepath 1 setgray fill grestore 0 setgray closepath } def
/DiaW { stroke [] 0 setdash vpt add M
  hpt neg vpt neg V hpt vpt neg V
  hpt vpt V hpt neg vpt V Opaque stroke } def
/BoxW { stroke [] 0 setdash exch hpt sub exch vpt add M
  0 vpt2 neg V hpt2 0 V 0 vpt2 V
  hpt2 neg 0 V Opaque stroke } def
/TriUW { stroke [] 0 setdash vpt 1.12 mul add M
  hpt neg vpt -1.62 mul V
  hpt 2 mul 0 V
  hpt neg vpt 1.62 mul V Opaque stroke } def
/TriDW { stroke [] 0 setdash vpt 1.12 mul sub M
  hpt neg vpt 1.62 mul V
  hpt 2 mul 0 V
  hpt neg vpt -1.62 mul V Opaque stroke } def
/PentW { stroke [] 0 setdash gsave
  translate 0 hpt M 4 {72 rotate 0 hpt L} repeat
  Opaque stroke grestore } def
/CircW { stroke [] 0 setdash 
  hpt 0 360 arc Opaque stroke } def
/BoxFill { gsave Rec 1 setgray fill grestore } def
end
}}%
\begin{picture}(1800,1296)(0,0)%
{\GNUPLOTspecial{"
gnudict begin
gsave
0 0 translate
0.100 0.100 scale
0 setgray
newpath
1.000 UL
LTb
400 300 M
63 0 V
1187 0 R
-63 0 V
400 479 M
63 0 V
1187 0 R
-63 0 V
400 658 M
63 0 V
1187 0 R
-63 0 V
400 838 M
63 0 V
1187 0 R
-63 0 V
400 1017 M
63 0 V
1187 0 R
-63 0 V
400 1196 M
63 0 V
1187 0 R
-63 0 V
400 300 M
0 63 V
0 833 R
0 -63 V
650 300 M
0 63 V
0 833 R
0 -63 V
900 300 M
0 63 V
0 833 R
0 -63 V
1150 300 M
0 63 V
0 833 R
0 -63 V
1400 300 M
0 63 V
0 833 R
0 -63 V
1650 300 M
0 63 V
0 833 R
0 -63 V
1.000 UL
LTb
400 300 M
1250 0 V
0 896 V
-1250 0 V
0 -896 V
1.000 UL
LT0
405 1032 M
5 -1 V
5 -2 V
5 -2 V
5 -2 V
5 -2 V
5 -2 V
5 -2 V
5 -1 V
5 -2 V
5 -2 V
5 -2 V
5 -2 V
5 -2 V
5 -2 V
5 -2 V
5 -2 V
5 -2 V
5 -2 V
5 -2 V
5 -2 V
5 -3 V
5 -2 V
5 -2 V
5 -2 V
5 -2 V
5 -2 V
5 -2 V
5 -2 V
6 -3 V
5 -2 V
5 -2 V
5 -2 V
5 -3 V
5 -2 V
5 -2 V
5 -2 V
5 -3 V
5 -2 V
5 -2 V
5 -3 V
5 -2 V
5 -2 V
5 -3 V
5 -2 V
5 -3 V
5 -2 V
5 -3 V
5 -2 V
5 -2 V
5 -3 V
5 -2 V
5 -3 V
5 -2 V
5 -3 V
5 -3 V
5 -2 V
5 -3 V
5 -2 V
5 -3 V
5 -3 V
5 -2 V
5 -3 V
5 -3 V
5 -2 V
5 -3 V
5 -3 V
5 -2 V
5 -3 V
5 -3 V
5 -3 V
5 -2 V
5 -3 V
5 -3 V
5 -3 V
5 -3 V
5 -3 V
5 -3 V
5 -2 V
5 -3 V
5 -3 V
5 -3 V
5 -3 V
5 -3 V
5 -3 V
5 -3 V
5 -3 V
5 -3 V
5 -3 V
6 -3 V
5 -4 V
5 -3 V
5 -3 V
5 -3 V
5 -3 V
5 -3 V
5 -3 V
5 -4 V
5 -3 V
5 -3 V
5 -3 V
5 -4 V
5 -3 V
5 -3 V
5 -4 V
5 -3 V
5 -3 V
5 -4 V
5 -3 V
5 -4 V
5 -3 V
5 -4 V
5 -3 V
5 -4 V
5 -3 V
5 -4 V
5 -3 V
5 -4 V
5 -3 V
5 -4 V
5 -4 V
5 -3 V
5 -4 V
5 -4 V
5 -3 V
5 -4 V
5 -4 V
5 -4 V
5 -4 V
5 -3 V
5 -4 V
5 -4 V
5 -4 V
5 -4 V
5 -4 V
5 -4 V
5 -4 V
5 -4 V
5 -4 V
5 -4 V
5 -4 V
5 -4 V
5 -5 V
5 -4 V
5 -4 V
5 -4 V
5 -5 V
5 -4 V
5 -4 V
6 -4 V
5 -5 V
5 -4 V
5 -5 V
5 -4 V
5 -5 V
5 -4 V
5 -5 V
5 -4 V
5 -5 V
5 -5 V
5 -4 V
5 -5 V
5 -5 V
5 -5 V
5 -4 V
5 -5 V
5 -5 V
5 -5 V
5 -5 V
5 -5 V
5 -5 V
5 -5 V
5 -5 V
5 -6 V
5 -5 V
5 -5 V
5 -5 V
5 -6 V
5 -5 V
5 -5 V
5 -6 V
5 -6 V
5 -5 V
5 -6 V
5 -5 V
5 -6 V
5 -6 V
5 -6 V
5 -6 V
5 -6 V
5 -6 V
5 -6 V
5 -6 V
5 -7 V
5 -6 V
5 -6 V
5 -7 V
5 -6 V
5 -7 V
5 -7 V
5 -7 V
5 -6 V
5 -8 V
5 -7 V
1 -1 V
1.000 UL
LT0
405 1036 M
5 2 V
5 2 V
5 1 V
5 2 V
5 2 V
5 2 V
5 1 V
5 2 V
5 2 V
5 1 V
5 2 V
5 2 V
5 1 V
5 2 V
5 1 V
5 2 V
5 1 V
5 2 V
5 2 V
5 1 V
5 2 V
5 1 V
5 2 V
5 1 V
5 2 V
5 1 V
5 1 V
5 2 V
6 1 V
5 2 V
5 1 V
5 1 V
5 2 V
5 1 V
5 1 V
5 2 V
5 1 V
5 1 V
5 1 V
5 2 V
5 1 V
5 1 V
5 1 V
5 2 V
5 1 V
5 1 V
5 1 V
5 1 V
5 1 V
5 1 V
5 2 V
5 1 V
5 1 V
5 1 V
5 1 V
5 1 V
5 1 V
5 1 V
5 1 V
5 1 V
5 1 V
5 1 V
5 1 V
5 1 V
5 0 V
5 1 V
5 1 V
5 1 V
5 1 V
5 1 V
5 0 V
5 1 V
5 1 V
5 1 V
5 1 V
5 0 V
5 1 V
5 1 V
5 0 V
5 1 V
5 1 V
5 0 V
5 1 V
5 0 V
5 1 V
5 1 V
5 0 V
5 1 V
6 0 V
5 1 V
5 0 V
5 1 V
5 0 V
5 1 V
5 0 V
5 0 V
5 1 V
5 0 V
5 0 V
5 1 V
5 0 V
5 0 V
5 1 V
5 0 V
5 0 V
5 0 V
5 1 V
5 0 V
5 0 V
5 0 V
5 0 V
5 0 V
5 0 V
5 0 V
5 0 V
5 1 V
5 0 V
5 0 V
5 -1 V
5 0 V
5 0 V
5 0 V
5 0 V
5 0 V
5 0 V
5 0 V
5 -1 V
5 0 V
5 0 V
5 0 V
5 -1 V
5 0 V
5 0 V
5 -1 V
5 0 V
5 0 V
5 -1 V
5 0 V
5 -1 V
5 0 V
5 -1 V
5 0 V
5 -1 V
5 -1 V
5 0 V
5 -1 V
5 -1 V
5 0 V
6 -1 V
5 -1 V
5 -1 V
5 -1 V
5 0 V
5 -1 V
5 -1 V
5 -1 V
5 -1 V
5 -1 V
5 -1 V
5 -1 V
5 -1 V
5 -2 V
5 -1 V
5 -1 V
5 -1 V
5 -2 V
5 -1 V
5 -1 V
5 -2 V
5 -1 V
5 -2 V
5 -1 V
5 -2 V
5 -1 V
5 -2 V
5 -2 V
5 -2 V
5 -1 V
5 -2 V
5 -2 V
5 -2 V
5 -2 V
5 -2 V
5 -2 V
5 -2 V
5 -3 V
5 -2 V
5 -2 V
5 -3 V
5 -2 V
5 -3 V
5 -2 V
5 -3 V
5 -3 V
5 -2 V
5 -3 V
5 -3 V
5 -3 V
5 -4 V
5 -3 V
5 -3 V
5 -4 V
5 -3 V
5 -4 V
5 -3 V
5 -4 V
5 -4 V
5 -4 V
6 -5 V
5 -4 V
5 -5 V
5 -4 V
5 -5 V
5 -5 V
5 -5 V
5 -6 V
5 -5 V
5 -6 V
5 -7 V
5 -6 V
5 -7 V
5 -7 V
5 -7 V
5 -8 V
5 -9 V
5 -9 V
5 -10 V
5 -10 V
5 -12 V
5 -13 V
5 -14 V
5 -17 V
5 -21 V
5 -30 V
1.000 UL
LT0
1424 300 M
4 6 V
5 7 V
5 8 V
5 7 V
5 8 V
6 8 V
5 8 V
5 8 V
5 8 V
5 9 V
5 8 V
5 9 V
5 9 V
5 9 V
5 10 V
5 10 V
5 10 V
5 10 V
5 11 V
5 11 V
5 11 V
5 12 V
5 13 V
5 13 V
5 15 V
5 15 V
5 16 V
5 18 V
5 21 V
5 25 V
5 33 V
0 113 V
1.000 UL
LT1
405 1020 M
5 2 V
5 2 V
5 2 V
5 1 V
5 2 V
5 2 V
5 1 V
5 2 V
5 2 V
5 1 V
5 2 V
5 2 V
5 1 V
5 2 V
5 2 V
5 1 V
5 2 V
5 1 V
5 2 V
5 1 V
5 2 V
5 1 V
5 2 V
5 1 V
5 2 V
5 1 V
5 2 V
5 1 V
6 1 V
5 2 V
5 1 V
5 1 V
5 2 V
5 1 V
5 1 V
5 2 V
5 1 V
5 1 V
5 2 V
5 1 V
5 1 V
5 1 V
5 1 V
5 2 V
5 1 V
5 1 V
5 1 V
5 1 V
5 1 V
5 1 V
5 2 V
5 1 V
5 1 V
5 1 V
5 1 V
5 1 V
5 1 V
5 1 V
5 1 V
5 1 V
5 1 V
5 0 V
5 1 V
5 1 V
5 1 V
5 1 V
5 1 V
5 1 V
5 0 V
5 1 V
5 1 V
5 1 V
5 1 V
5 0 V
5 1 V
5 1 V
5 0 V
5 1 V
5 1 V
5 0 V
5 1 V
5 1 V
5 0 V
5 1 V
5 0 V
5 1 V
5 0 V
5 1 V
6 0 V
5 1 V
5 0 V
5 1 V
5 0 V
5 1 V
5 0 V
5 0 V
5 1 V
5 0 V
5 0 V
5 1 V
5 0 V
5 0 V
5 0 V
5 1 V
5 0 V
5 0 V
5 0 V
5 0 V
5 0 V
5 0 V
5 1 V
5 0 V
5 0 V
5 0 V
5 0 V
5 0 V
5 0 V
5 0 V
5 -1 V
5 0 V
5 0 V
5 0 V
5 0 V
5 0 V
5 -1 V
5 0 V
5 0 V
5 0 V
5 -1 V
5 0 V
5 0 V
5 -1 V
5 0 V
5 -1 V
5 0 V
5 -1 V
5 0 V
5 -1 V
5 0 V
5 -1 V
5 0 V
5 -1 V
5 -1 V
5 0 V
5 -1 V
5 -1 V
5 -1 V
5 -1 V
6 0 V
5 -1 V
5 -1 V
5 -1 V
5 -1 V
5 -1 V
5 -1 V
5 -1 V
5 -1 V
5 -1 V
5 -2 V
5 -1 V
5 -1 V
5 -1 V
5 -2 V
5 -1 V
5 -2 V
5 -1 V
5 -1 V
5 -2 V
5 -2 V
5 -1 V
5 -2 V
5 -2 V
5 -1 V
5 -2 V
5 -2 V
5 -2 V
5 -2 V
5 -2 V
5 -2 V
5 -2 V
5 -2 V
5 -2 V
5 -3 V
5 -2 V
5 -2 V
5 -3 V
5 -2 V
5 -3 V
5 -3 V
5 -3 V
5 -2 V
5 -3 V
5 -3 V
5 -3 V
5 -4 V
5 -3 V
5 -3 V
5 -4 V
5 -3 V
5 -4 V
5 -4 V
5 -4 V
5 -4 V
5 -4 V
5 -4 V
5 -5 V
5 -4 V
5 -5 V
6 -5 V
5 -5 V
5 -5 V
5 -6 V
5 -6 V
5 -6 V
5 -6 V
5 -7 V
5 -7 V
5 -7 V
5 -8 V
5 -8 V
5 -9 V
5 -10 V
5 -10 V
5 -12 V
5 -12 V
5 -14 V
5 -17 V
5 -21 V
5 -29 V
1.000 UL
LT1
1402 300 M
1 1 V
5 8 V
5 7 V
5 8 V
5 7 V
5 8 V
5 8 V
5 8 V
5 8 V
5 8 V
6 9 V
5 9 V
5 9 V
5 9 V
5 9 V
5 10 V
5 10 V
5 10 V
5 11 V
5 11 V
5 11 V
5 12 V
5 13 V
5 13 V
5 14 V
5 15 V
5 16 V
5 18 V
5 20 V
5 24 V
5 33 V
0 113 V
1.000 UL
LT1
405 1017 M
5 -2 V
5 -2 V
5 -2 V
5 -2 V
5 -2 V
5 -1 V
5 -2 V
5 -2 V
5 -2 V
5 -2 V
5 -2 V
5 -2 V
5 -2 V
5 -2 V
5 -2 V
5 -2 V
5 -2 V
5 -2 V
5 -2 V
5 -2 V
5 -2 V
5 -2 V
5 -3 V
5 -2 V
5 -2 V
5 -2 V
5 -2 V
5 -2 V
6 -3 V
5 -2 V
5 -2 V
5 -2 V
5 -2 V
5 -3 V
5 -2 V
5 -2 V
5 -3 V
5 -2 V
5 -2 V
5 -3 V
5 -2 V
5 -2 V
5 -3 V
5 -2 V
5 -3 V
5 -2 V
5 -3 V
5 -2 V
5 -2 V
5 -3 V
5 -2 V
5 -3 V
5 -3 V
5 -2 V
5 -3 V
5 -2 V
5 -3 V
5 -2 V
5 -3 V
5 -3 V
5 -2 V
5 -3 V
5 -3 V
5 -2 V
5 -3 V
5 -3 V
5 -3 V
5 -2 V
5 -3 V
5 -3 V
5 -3 V
5 -3 V
5 -3 V
5 -2 V
5 -3 V
5 -3 V
5 -3 V
5 -3 V
5 -3 V
5 -3 V
5 -3 V
5 -3 V
5 -3 V
5 -3 V
5 -3 V
5 -3 V
5 -3 V
5 -3 V
6 -3 V
5 -3 V
5 -4 V
5 -3 V
5 -3 V
5 -3 V
5 -3 V
5 -3 V
5 -4 V
5 -3 V
5 -3 V
5 -4 V
5 -3 V
5 -3 V
5 -4 V
5 -3 V
5 -3 V
5 -4 V
5 -3 V
5 -4 V
5 -3 V
5 -4 V
5 -3 V
5 -4 V
5 -3 V
5 -4 V
5 -3 V
5 -4 V
5 -4 V
5 -3 V
5 -4 V
5 -4 V
5 -4 V
5 -3 V
5 -4 V
5 -4 V
5 -4 V
5 -3 V
5 -4 V
5 -4 V
5 -4 V
5 -4 V
5 -4 V
5 -4 V
5 -4 V
5 -4 V
5 -4 V
5 -4 V
5 -4 V
5 -4 V
5 -5 V
5 -4 V
5 -4 V
5 -4 V
5 -5 V
5 -4 V
5 -4 V
5 -5 V
5 -4 V
5 -4 V
6 -5 V
5 -4 V
5 -5 V
5 -4 V
5 -5 V
5 -4 V
5 -5 V
5 -5 V
5 -5 V
5 -4 V
5 -5 V
5 -5 V
5 -5 V
5 -5 V
5 -5 V
5 -4 V
5 -5 V
5 -6 V
5 -5 V
5 -5 V
5 -5 V
5 -5 V
5 -5 V
5 -6 V
5 -5 V
5 -5 V
5 -6 V
5 -5 V
5 -6 V
5 -6 V
5 -5 V
5 -6 V
5 -6 V
5 -5 V
5 -6 V
5 -6 V
5 -6 V
5 -6 V
5 -7 V
5 -6 V
5 -6 V
5 -6 V
5 -7 V
5 -6 V
5 -7 V
5 -7 V
5 -6 V
5 -7 V
5 -7 V
5 -7 V
4 -6 V
1.000 UL
LT1
stroke
grestore
end
showpage
}}%
\put(1025,50){\makebox(0,0){$p_{ij}$  [fm $^ {-1}$]}}%
\put(100,748){%
\special{ps: gsave currentpoint currentpoint translate
270 rotate neg exch neg exch translate}%
\makebox(0,0)[b]{\shortstack{$p_{jk}$  [fm $^ {-1}$]}}%
\special{ps: currentpoint grestore moveto}%
}%
\put(1650,200){\makebox(0,0){2.5}}%
\put(1400,200){\makebox(0,0){2}}%
\put(1150,200){\makebox(0,0){1.5}}%
\put(900,200){\makebox(0,0){1}}%
\put(650,200){\makebox(0,0){0.5}}%
\put(400,200){\makebox(0,0){0}}%
\put(350,1196){\makebox(0,0)[r]{2.5}}%
\put(350,1017){\makebox(0,0)[r]{2}}%
\put(350,838){\makebox(0,0)[r]{1.5}}%
\put(350,658){\makebox(0,0)[r]{1}}%
\put(350,479){\makebox(0,0)[r]{0.5}}%
\put(350,300){\makebox(0,0)[r]{0}}%
\end{picture}%
\endgroup
 

%% file: fig2.tex
\begingroup%
  \makeatletter%
  \newcommand{\GNUPLOTspecial}{%
    \@sanitize\catcode`\%=14\relax\special}%
  \setlength{\unitlength}{0.1bp}%
{\GNUPLOTspecial{!
/gnudict 256 dict def
gnudict begin
/Color false def
/Solid false def
/gnulinewidth 5.000 def
/userlinewidth gnulinewidth def
/vshift -33 def
/dl {10 mul} def
/hpt_ 31.5 def
/vpt_ 31.5 def
/hpt hpt_ def
/vpt vpt_ def
/M {moveto} bind def
/L {lineto} bind def
/R {rmoveto} bind def
/V {rlineto} bind def
/vpt2 vpt 2 mul def
/hpt2 hpt 2 mul def
/Lshow { currentpoint stroke M
  0 vshift R show } def
/Rshow { currentpoint stroke M
  dup stringwidth pop neg vshift R show } def
/Cshow { currentpoint stroke M
  dup stringwidth pop -2 div vshift R show } def
/UP { dup vpt_ mul /vpt exch def hpt_ mul /hpt exch def
  /hpt2 hpt 2 mul def /vpt2 vpt 2 mul def } def
/DL { Color {setrgbcolor Solid {pop []} if 0 setdash }
 {pop pop pop Solid {pop []} if 0 setdash} ifelse } def
/BL { stroke gnulinewidth 2 mul setlinewidth } def
/AL { stroke gnulinewidth 2 div setlinewidth } def
/UL { gnulinewidth mul /userlinewidth exch def } def
/PL { stroke userlinewidth setlinewidth } def
/LTb { BL [] 0 0 0 DL } def
/LTa { AL [1 dl 2 dl] 0 setdash 0 0 0 setrgbcolor } def
/LT0 { PL [] 1 0 0 DL } def
/LT1 { PL [4 dl 2 dl] 0 1 0 DL } def
/LT2 { PL [2 dl 3 dl] 0 0 1 DL } def
/LT3 { PL [1 dl 1.5 dl] 1 0 1 DL } def
/LT4 { PL [5 dl 2 dl 1 dl 2 dl] 0 1 1 DL } def
/LT5 { PL [4 dl 3 dl 1 dl 3 dl] 1 1 0 DL } def
/LT6 { PL [2 dl 2 dl 2 dl 4 dl] 0 0 0 DL } def
/LT7 { PL [2 dl 2 dl 2 dl 2 dl 2 dl 4 dl] 1 0.3 0 DL } def
/LT8 { PL [2 dl 2 dl 2 dl 2 dl 2 dl 2 dl 2 dl 4 dl] 0.5 0.5 0.5 DL } def
/Pnt { stroke [] 0 setdash
   gsave 1 setlinecap M 0 0 V stroke grestore } def
/Dia { stroke [] 0 setdash 2 copy vpt add M
  hpt neg vpt neg V hpt vpt neg V
  hpt vpt V hpt neg vpt V closepath stroke
  Pnt } def
/Pls { stroke [] 0 setdash vpt sub M 0 vpt2 V
  currentpoint stroke M
  hpt neg vpt neg R hpt2 0 V stroke
  } def
/Box { stroke [] 0 setdash 2 copy exch hpt sub exch vpt add M
  0 vpt2 neg V hpt2 0 V 0 vpt2 V
  hpt2 neg 0 V closepath stroke
  Pnt } def
/Crs { stroke [] 0 setdash exch hpt sub exch vpt add M
  hpt2 vpt2 neg V currentpoint stroke M
  hpt2 neg 0 R hpt2 vpt2 V stroke } def
/TriU { stroke [] 0 setdash 2 copy vpt 1.12 mul add M
  hpt neg vpt -1.62 mul V
  hpt 2 mul 0 V
  hpt neg vpt 1.62 mul V closepath stroke
  Pnt  } def
/Star { 2 copy Pls Crs } def
/BoxF { stroke [] 0 setdash exch hpt sub exch vpt add M
  0 vpt2 neg V  hpt2 0 V  0 vpt2 V
  hpt2 neg 0 V  closepath fill } def
/TriUF { stroke [] 0 setdash vpt 1.12 mul add M
  hpt neg vpt -1.62 mul V
  hpt 2 mul 0 V
  hpt neg vpt 1.62 mul V closepath fill } def
/TriD { stroke [] 0 setdash 2 copy vpt 1.12 mul sub M
  hpt neg vpt 1.62 mul V
  hpt 2 mul 0 V
  hpt neg vpt -1.62 mul V closepath stroke
  Pnt  } def
/TriDF { stroke [] 0 setdash vpt 1.12 mul sub M
  hpt neg vpt 1.62 mul V
  hpt 2 mul 0 V
  hpt neg vpt -1.62 mul V closepath fill} def
/DiaF { stroke [] 0 setdash vpt add M
  hpt neg vpt neg V hpt vpt neg V
  hpt vpt V hpt neg vpt V closepath fill } def
/Pent { stroke [] 0 setdash 2 copy gsave
  translate 0 hpt M 4 {72 rotate 0 hpt L} repeat
  closepath stroke grestore Pnt } def
/PentF { stroke [] 0 setdash gsave
  translate 0 hpt M 4 {72 rotate 0 hpt L} repeat
  closepath fill grestore } def
/Circle { stroke [] 0 setdash 2 copy
  hpt 0 360 arc stroke Pnt } def
/CircleF { stroke [] 0 setdash hpt 0 360 arc fill } def
/C0 { BL [] 0 setdash 2 copy moveto vpt 90 450  arc } bind def
/C1 { BL [] 0 setdash 2 copy        moveto
       2 copy  vpt 0 90 arc closepath fill
               vpt 0 360 arc closepath } bind def
/C2 { BL [] 0 setdash 2 copy moveto
       2 copy  vpt 90 180 arc closepath fill
               vpt 0 360 arc closepath } bind def
/C3 { BL [] 0 setdash 2 copy moveto
       2 copy  vpt 0 180 arc closepath fill
               vpt 0 360 arc closepath } bind def
/C4 { BL [] 0 setdash 2 copy moveto
       2 copy  vpt 180 270 arc closepath fill
               vpt 0 360 arc closepath } bind def
/C5 { BL [] 0 setdash 2 copy moveto
       2 copy  vpt 0 90 arc
       2 copy moveto
       2 copy  vpt 180 270 arc closepath fill
               vpt 0 360 arc } bind def
/C6 { BL [] 0 setdash 2 copy moveto
      2 copy  vpt 90 270 arc closepath fill
              vpt 0 360 arc closepath } bind def
/C7 { BL [] 0 setdash 2 copy moveto
      2 copy  vpt 0 270 arc closepath fill
              vpt 0 360 arc closepath } bind def
/C8 { BL [] 0 setdash 2 copy moveto
      2 copy vpt 270 360 arc closepath fill
              vpt 0 360 arc closepath } bind def
/C9 { BL [] 0 setdash 2 copy moveto
      2 copy  vpt 270 450 arc closepath fill
              vpt 0 360 arc closepath } bind def
/C10 { BL [] 0 setdash 2 copy 2 copy moveto vpt 270 360 arc closepath fill
       2 copy moveto
       2 copy vpt 90 180 arc closepath fill
               vpt 0 360 arc closepath } bind def
/C11 { BL [] 0 setdash 2 copy moveto
       2 copy  vpt 0 180 arc closepath fill
       2 copy moveto
       2 copy  vpt 270 360 arc closepath fill
               vpt 0 360 arc closepath } bind def
/C12 { BL [] 0 setdash 2 copy moveto
       2 copy  vpt 180 360 arc closepath fill
               vpt 0 360 arc closepath } bind def
/C13 { BL [] 0 setdash  2 copy moveto
       2 copy  vpt 0 90 arc closepath fill
       2 copy moveto
       2 copy  vpt 180 360 arc closepath fill
               vpt 0 360 arc closepath } bind def
/C14 { BL [] 0 setdash 2 copy moveto
       2 copy  vpt 90 360 arc closepath fill
               vpt 0 360 arc } bind def
/C15 { BL [] 0 setdash 2 copy vpt 0 360 arc closepath fill
               vpt 0 360 arc closepath } bind def
/Rec   { newpath 4 2 roll moveto 1 index 0 rlineto 0 exch rlineto
       neg 0 rlineto closepath } bind def
/Square { dup Rec } bind def
/Bsquare { vpt sub exch vpt sub exch vpt2 Square } bind def
/S0 { BL [] 0 setdash 2 copy moveto 0 vpt rlineto BL Bsquare } bind def
/S1 { BL [] 0 setdash 2 copy vpt Square fill Bsquare } bind def
/S2 { BL [] 0 setdash 2 copy exch vpt sub exch vpt Square fill Bsquare } bind def
/S3 { BL [] 0 setdash 2 copy exch vpt sub exch vpt2 vpt Rec fill Bsquare } bind def
/S4 { BL [] 0 setdash 2 copy exch vpt sub exch vpt sub vpt Square fill Bsquare } bind def
/S5 { BL [] 0 setdash 2 copy 2 copy vpt Square fill
       exch vpt sub exch vpt sub vpt Square fill Bsquare } bind def
/S6 { BL [] 0 setdash 2 copy exch vpt sub exch vpt sub vpt vpt2 Rec fill Bsquare } bind def
/S7 { BL [] 0 setdash 2 copy exch vpt sub exch vpt sub vpt vpt2 Rec fill
       2 copy vpt Square fill
       Bsquare } bind def
/S8 { BL [] 0 setdash 2 copy vpt sub vpt Square fill Bsquare } bind def
/S9 { BL [] 0 setdash 2 copy vpt sub vpt vpt2 Rec fill Bsquare } bind def
/S10 { BL [] 0 setdash 2 copy vpt sub vpt Square fill 2 copy exch vpt sub exch vpt Square fill
       Bsquare } bind def
/S11 { BL [] 0 setdash 2 copy vpt sub vpt Square fill 2 copy exch vpt sub exch vpt2 vpt Rec fill
       Bsquare } bind def
/S12 { BL [] 0 setdash 2 copy exch vpt sub exch vpt sub vpt2 vpt Rec fill Bsquare } bind def
/S13 { BL [] 0 setdash 2 copy exch vpt sub exch vpt sub vpt2 vpt Rec fill
       2 copy vpt Square fill Bsquare } bind def
/S14 { BL [] 0 setdash 2 copy exch vpt sub exch vpt sub vpt2 vpt Rec fill
       2 copy exch vpt sub exch vpt Square fill Bsquare } bind def
/S15 { BL [] 0 setdash 2 copy Bsquare fill Bsquare } bind def
/D0 { gsave translate 45 rotate 0 0 S0 stroke grestore } bind def
/D1 { gsave translate 45 rotate 0 0 S1 stroke grestore } bind def
/D2 { gsave translate 45 rotate 0 0 S2 stroke grestore } bind def
/D3 { gsave translate 45 rotate 0 0 S3 stroke grestore } bind def
/D4 { gsave translate 45 rotate 0 0 S4 stroke grestore } bind def
/D5 { gsave translate 45 rotate 0 0 S5 stroke grestore } bind def
/D6 { gsave translate 45 rotate 0 0 S6 stroke grestore } bind def
/D7 { gsave translate 45 rotate 0 0 S7 stroke grestore } bind def
/D8 { gsave translate 45 rotate 0 0 S8 stroke grestore } bind def
/D9 { gsave translate 45 rotate 0 0 S9 stroke grestore } bind def
/D10 { gsave translate 45 rotate 0 0 S10 stroke grestore } bind def
/D11 { gsave translate 45 rotate 0 0 S11 stroke grestore } bind def
/D12 { gsave translate 45 rotate 0 0 S12 stroke grestore } bind def
/D13 { gsave translate 45 rotate 0 0 S13 stroke grestore } bind def
/D14 { gsave translate 45 rotate 0 0 S14 stroke grestore } bind def
/D15 { gsave translate 45 rotate 0 0 S15 stroke grestore } bind def
/DiaE { stroke [] 0 setdash vpt add M
  hpt neg vpt neg V hpt vpt neg V
  hpt vpt V hpt neg vpt V closepath stroke } def
/BoxE { stroke [] 0 setdash exch hpt sub exch vpt add M
  0 vpt2 neg V hpt2 0 V 0 vpt2 V
  hpt2 neg 0 V closepath stroke } def
/TriUE { stroke [] 0 setdash vpt 1.12 mul add M
  hpt neg vpt -1.62 mul V
  hpt 2 mul 0 V
  hpt neg vpt 1.62 mul V closepath stroke } def
/TriDE { stroke [] 0 setdash vpt 1.12 mul sub M
  hpt neg vpt 1.62 mul V
  hpt 2 mul 0 V
  hpt neg vpt -1.62 mul V closepath stroke } def
/PentE { stroke [] 0 setdash gsave
  translate 0 hpt M 4 {72 rotate 0 hpt L} repeat
  closepath stroke grestore } def
/CircE { stroke [] 0 setdash 
  hpt 0 360 arc stroke } def
/Opaque { gsave closepath 1 setgray fill grestore 0 setgray closepath } def
/DiaW { stroke [] 0 setdash vpt add M
  hpt neg vpt neg V hpt vpt neg V
  hpt vpt V hpt neg vpt V Opaque stroke } def
/BoxW { stroke [] 0 setdash exch hpt sub exch vpt add M
  0 vpt2 neg V hpt2 0 V 0 vpt2 V
  hpt2 neg 0 V Opaque stroke } def
/TriUW { stroke [] 0 setdash vpt 1.12 mul add M
  hpt neg vpt -1.62 mul V
  hpt 2 mul 0 V
  hpt neg vpt 1.62 mul V Opaque stroke } def
/TriDW { stroke [] 0 setdash vpt 1.12 mul sub M
  hpt neg vpt 1.62 mul V
  hpt 2 mul 0 V
  hpt neg vpt -1.62 mul V Opaque stroke } def
/PentW { stroke [] 0 setdash gsave
  translate 0 hpt M 4 {72 rotate 0 hpt L} repeat
  Opaque stroke grestore } def
/CircW { stroke [] 0 setdash 
  hpt 0 360 arc Opaque stroke } def
/BoxFill { gsave Rec 1 setgray fill grestore } def
end
}}%
\begin{picture}(2339,1296)(0,0)%
{\GNUPLOTspecial{"
gnudict begin
gsave
0 0 translate
0.100 0.100 scale
0 setgray
newpath
1.000 UL
LTb
400 300 M
63 0 V
1826 0 R
-63 0 V
400 479 M
63 0 V
1826 0 R
-63 0 V
400 658 M
63 0 V
1826 0 R
-63 0 V
400 838 M
63 0 V
1826 0 R
-63 0 V
400 1017 M
63 0 V
1826 0 R
-63 0 V
400 1196 M
63 0 V
1826 0 R
-63 0 V
400 300 M
0 63 V
0 833 R
0 -63 V
778 300 M
0 63 V
0 833 R
0 -63 V
1156 300 M
0 63 V
0 833 R
0 -63 V
1533 300 M
0 63 V
0 833 R
0 -63 V
1911 300 M
0 63 V
0 833 R
0 -63 V
2289 300 M
0 63 V
0 833 R
0 -63 V
1.000 UL
LTb
400 300 M
1889 0 V
0 896 V
-1889 0 V
0 -896 V
1.000 UL
LT0
1926 1083 M
263 0 V
438 347 M
38 25 V
37 121 V
38 169 V
38 158 V
38 122 V
37 78 V
38 42 V
38 13 V
38 -7 V
38 -21 V
37 -28 V
38 -32 V
38 -35 V
38 -35 V
37 -35 V
38 -33 V
38 -32 V
38 -30 V
38 -29 V
37 -27 V
38 -25 V
38 -23 V
38 -22 V
38 -20 V
37 -19 V
38 -18 V
38 -16 V
38 -16 V
37 -14 V
38 -14 V
38 -12 V
38 -12 V
38 -11 V
37 -10 V
38 -10 V
38 -9 V
38 -9 V
37 -8 V
38 -7 V
38 -8 V
38 -6 V
38 -7 V
37 -6 V
38 -5 V
38 -6 V
38 -5 V
37 -5 V
38 -4 V
38 -5 V
1.000 UL
LT0
438 310 M
38 21 V
37 58 V
38 71 V
38 60 V
38 43 V
37 25 V
38 10 V
38 -3 V
38 -10 V
38 -14 V
37 -17 V
38 -18 V
38 -18 V
38 -18 V
37 -16 V
38 -16 V
38 -14 V
38 -13 V
38 -12 V
37 -11 V
38 -10 V
38 -9 V
38 -8 V
38 -8 V
37 -6 V
38 -7 V
38 -5 V
38 -5 V
37 -5 V
38 -4 V
38 -4 V
38 -3 V
38 -4 V
37 -2 V
38 -3 V
38 -3 V
38 -2 V
37 -2 V
38 -2 V
38 -2 V
38 -1 V
38 -2 V
37 -1 V
38 -1 V
38 -2 V
38 -1 V
37 -1 V
38 -1 V
38 0 V
1.000 UL
LT3
1926 983 M
263 0 V
438 464 M
38 38 V
37 67 V
38 96 V
38 111 V
38 108 V
37 86 V
38 58 V
38 28 V
38 5 V
38 -13 V
37 -25 V
38 -32 V
38 -34 V
38 -36 V
37 -35 V
38 -34 V
38 -32 V
38 -31 V
38 -29 V
37 -27 V
38 -26 V
38 -23 V
38 -22 V
38 -21 V
37 -19 V
38 -17 V
38 -17 V
38 -15 V
37 -15 V
38 -13 V
38 -13 V
38 -12 V
38 -11 V
37 -10 V
38 -10 V
38 -9 V
38 -9 V
37 -8 V
38 -7 V
38 -7 V
38 -7 V
38 -7 V
37 -6 V
38 -5 V
38 -6 V
38 -5 V
37 -5 V
38 -4 V
38 -5 V
1.000 UL
LT3
438 309 M
38 21 V
37 34 V
38 46 V
38 52 V
38 49 V
37 37 V
38 24 V
38 10 V
38 -1 V
38 -10 V
37 -14 V
38 -17 V
38 -17 V
38 -18 V
37 -17 V
38 -16 V
38 -15 V
38 -14 V
38 -12 V
37 -11 V
38 -10 V
38 -10 V
38 -8 V
38 -8 V
37 -7 V
38 -6 V
38 -6 V
38 -5 V
37 -4 V
38 -5 V
38 -3 V
38 -4 V
38 -3 V
37 -3 V
38 -3 V
38 -2 V
38 -3 V
37 -2 V
38 -2 V
38 -2 V
38 -1 V
38 -2 V
37 -1 V
38 -1 V
38 -1 V
38 -2 V
37 -1 V
38 -1 V
38 0 V
stroke
grestore
end
showpage
}}%
\put(1876,983){\makebox(0,0)[r]{Nonrelativistic}}%
\put(1876,1083){\makebox(0,0)[r]{Relativistic}}%
\put(1156,479){\makebox(0,0)[l]{D-wave}}%
\put(1345,748){\makebox(0,0)[l]{S-wave}}%
\put(1344,50){\makebox(0,0){r [ fm ]}}%
\put(100,748){%
\special{ps: gsave currentpoint currentpoint translate
270 rotate neg exch neg exch translate}%
\makebox(0,0)[b]{\shortstack{$\psi,\phi$ [fm$^{-3/2}$] }}%
\special{ps: currentpoint grestore moveto}%
}%
\put(2289,200){\makebox(0,0){5}}%
\put(1911,200){\makebox(0,0){4}}%
\put(1533,200){\makebox(0,0){3}}%
\put(1156,200){\makebox(0,0){2}}%
\put(778,200){\makebox(0,0){1}}%
\put(400,200){\makebox(0,0){0}}%
\put(350,1196){\makebox(0,0)[r]{0.5}}%
\put(350,1017){\makebox(0,0)[r]{0.4}}%
\put(350,838){\makebox(0,0)[r]{0.3}}%
\put(350,658){\makebox(0,0)[r]{0.2}}%
\put(350,479){\makebox(0,0)[r]{0.1}}%
\put(350,300){\makebox(0,0)[r]{0}}%
\end{picture}%
\endgroup
 

%% file: fig3.tex
\begingroup%
  \makeatletter%
  \newcommand{\GNUPLOTspecial}{%
    \@sanitize\catcode`\%=14\relax\special}%
  \setlength{\unitlength}{0.1bp}%
{\GNUPLOTspecial{!
/gnudict 256 dict def
gnudict begin
/Color false def
/Solid false def
/gnulinewidth 5.000 def
/userlinewidth gnulinewidth def
/vshift -33 def
/dl {10 mul} def
/hpt_ 31.5 def
/vpt_ 31.5 def
/hpt hpt_ def
/vpt vpt_ def
/M {moveto} bind def
/L {lineto} bind def
/R {rmoveto} bind def
/V {rlineto} bind def
/vpt2 vpt 2 mul def
/hpt2 hpt 2 mul def
/Lshow { currentpoint stroke M
  0 vshift R show } def
/Rshow { currentpoint stroke M
  dup stringwidth pop neg vshift R show } def
/Cshow { currentpoint stroke M
  dup stringwidth pop -2 div vshift R show } def
/UP { dup vpt_ mul /vpt exch def hpt_ mul /hpt exch def
  /hpt2 hpt 2 mul def /vpt2 vpt 2 mul def } def
/DL { Color {setrgbcolor Solid {pop []} if 0 setdash }
 {pop pop pop Solid {pop []} if 0 setdash} ifelse } def
/BL { stroke gnulinewidth 2 mul setlinewidth } def
/AL { stroke gnulinewidth 2 div setlinewidth } def
/UL { gnulinewidth mul /userlinewidth exch def } def
/PL { stroke userlinewidth setlinewidth } def
/LTb { BL [] 0 0 0 DL } def
/LTa { AL [1 dl 2 dl] 0 setdash 0 0 0 setrgbcolor } def
/LT0 { PL [] 1 0 0 DL } def
/LT1 { PL [4 dl 2 dl] 0 1 0 DL } def
/LT2 { PL [2 dl 3 dl] 0 0 1 DL } def
/LT3 { PL [1 dl 1.5 dl] 1 0 1 DL } def
/LT4 { PL [5 dl 2 dl 1 dl 2 dl] 0 1 1 DL } def
/LT5 { PL [4 dl 3 dl 1 dl 3 dl] 1 1 0 DL } def
/LT6 { PL [2 dl 2 dl 2 dl 4 dl] 0 0 0 DL } def
/LT7 { PL [2 dl 2 dl 2 dl 2 dl 2 dl 4 dl] 1 0.3 0 DL } def
/LT8 { PL [2 dl 2 dl 2 dl 2 dl 2 dl 2 dl 2 dl 4 dl] 0.5 0.5 0.5 DL } def
/Pnt { stroke [] 0 setdash
   gsave 1 setlinecap M 0 0 V stroke grestore } def
/Dia { stroke [] 0 setdash 2 copy vpt add M
  hpt neg vpt neg V hpt vpt neg V
  hpt vpt V hpt neg vpt V closepath stroke
  Pnt } def
/Pls { stroke [] 0 setdash vpt sub M 0 vpt2 V
  currentpoint stroke M
  hpt neg vpt neg R hpt2 0 V stroke
  } def
/Box { stroke [] 0 setdash 2 copy exch hpt sub exch vpt add M
  0 vpt2 neg V hpt2 0 V 0 vpt2 V
  hpt2 neg 0 V closepath stroke
  Pnt } def
/Crs { stroke [] 0 setdash exch hpt sub exch vpt add M
  hpt2 vpt2 neg V currentpoint stroke M
  hpt2 neg 0 R hpt2 vpt2 V stroke } def
/TriU { stroke [] 0 setdash 2 copy vpt 1.12 mul add M
  hpt neg vpt -1.62 mul V
  hpt 2 mul 0 V
  hpt neg vpt 1.62 mul V closepath stroke
  Pnt  } def
/Star { 2 copy Pls Crs } def
/BoxF { stroke [] 0 setdash exch hpt sub exch vpt add M
  0 vpt2 neg V  hpt2 0 V  0 vpt2 V
  hpt2 neg 0 V  closepath fill } def
/TriUF { stroke [] 0 setdash vpt 1.12 mul add M
  hpt neg vpt -1.62 mul V
  hpt 2 mul 0 V
  hpt neg vpt 1.62 mul V closepath fill } def
/TriD { stroke [] 0 setdash 2 copy vpt 1.12 mul sub M
  hpt neg vpt 1.62 mul V
  hpt 2 mul 0 V
  hpt neg vpt -1.62 mul V closepath stroke
  Pnt  } def
/TriDF { stroke [] 0 setdash vpt 1.12 mul sub M
  hpt neg vpt 1.62 mul V
  hpt 2 mul 0 V
  hpt neg vpt -1.62 mul V closepath fill} def
/DiaF { stroke [] 0 setdash vpt add M
  hpt neg vpt neg V hpt vpt neg V
  hpt vpt V hpt neg vpt V closepath fill } def
/Pent { stroke [] 0 setdash 2 copy gsave
  translate 0 hpt M 4 {72 rotate 0 hpt L} repeat
  closepath stroke grestore Pnt } def
/PentF { stroke [] 0 setdash gsave
  translate 0 hpt M 4 {72 rotate 0 hpt L} repeat
  closepath fill grestore } def
/Circle { stroke [] 0 setdash 2 copy
  hpt 0 360 arc stroke Pnt } def
/CircleF { stroke [] 0 setdash hpt 0 360 arc fill } def
/C0 { BL [] 0 setdash 2 copy moveto vpt 90 450  arc } bind def
/C1 { BL [] 0 setdash 2 copy        moveto
       2 copy  vpt 0 90 arc closepath fill
               vpt 0 360 arc closepath } bind def
/C2 { BL [] 0 setdash 2 copy moveto
       2 copy  vpt 90 180 arc closepath fill
               vpt 0 360 arc closepath } bind def
/C3 { BL [] 0 setdash 2 copy moveto
       2 copy  vpt 0 180 arc closepath fill
               vpt 0 360 arc closepath } bind def
/C4 { BL [] 0 setdash 2 copy moveto
       2 copy  vpt 180 270 arc closepath fill
               vpt 0 360 arc closepath } bind def
/C5 { BL [] 0 setdash 2 copy moveto
       2 copy  vpt 0 90 arc
       2 copy moveto
       2 copy  vpt 180 270 arc closepath fill
               vpt 0 360 arc } bind def
/C6 { BL [] 0 setdash 2 copy moveto
      2 copy  vpt 90 270 arc closepath fill
              vpt 0 360 arc closepath } bind def
/C7 { BL [] 0 setdash 2 copy moveto
      2 copy  vpt 0 270 arc closepath fill
              vpt 0 360 arc closepath } bind def
/C8 { BL [] 0 setdash 2 copy moveto
      2 copy vpt 270 360 arc closepath fill
              vpt 0 360 arc closepath } bind def
/C9 { BL [] 0 setdash 2 copy moveto
      2 copy  vpt 270 450 arc closepath fill
              vpt 0 360 arc closepath } bind def
/C10 { BL [] 0 setdash 2 copy 2 copy moveto vpt 270 360 arc closepath fill
       2 copy moveto
       2 copy vpt 90 180 arc closepath fill
               vpt 0 360 arc closepath } bind def
/C11 { BL [] 0 setdash 2 copy moveto
       2 copy  vpt 0 180 arc closepath fill
       2 copy moveto
       2 copy  vpt 270 360 arc closepath fill
               vpt 0 360 arc closepath } bind def
/C12 { BL [] 0 setdash 2 copy moveto
       2 copy  vpt 180 360 arc closepath fill
               vpt 0 360 arc closepath } bind def
/C13 { BL [] 0 setdash  2 copy moveto
       2 copy  vpt 0 90 arc closepath fill
       2 copy moveto
       2 copy  vpt 180 360 arc closepath fill
               vpt 0 360 arc closepath } bind def
/C14 { BL [] 0 setdash 2 copy moveto
       2 copy  vpt 90 360 arc closepath fill
               vpt 0 360 arc } bind def
/C15 { BL [] 0 setdash 2 copy vpt 0 360 arc closepath fill
               vpt 0 360 arc closepath } bind def
/Rec   { newpath 4 2 roll moveto 1 index 0 rlineto 0 exch rlineto
       neg 0 rlineto closepath } bind def
/Square { dup Rec } bind def
/Bsquare { vpt sub exch vpt sub exch vpt2 Square } bind def
/S0 { BL [] 0 setdash 2 copy moveto 0 vpt rlineto BL Bsquare } bind def
/S1 { BL [] 0 setdash 2 copy vpt Square fill Bsquare } bind def
/S2 { BL [] 0 setdash 2 copy exch vpt sub exch vpt Square fill Bsquare } bind def
/S3 { BL [] 0 setdash 2 copy exch vpt sub exch vpt2 vpt Rec fill Bsquare } bind def
/S4 { BL [] 0 setdash 2 copy exch vpt sub exch vpt sub vpt Square fill Bsquare } bind def
/S5 { BL [] 0 setdash 2 copy 2 copy vpt Square fill
       exch vpt sub exch vpt sub vpt Square fill Bsquare } bind def
/S6 { BL [] 0 setdash 2 copy exch vpt sub exch vpt sub vpt vpt2 Rec fill Bsquare } bind def
/S7 { BL [] 0 setdash 2 copy exch vpt sub exch vpt sub vpt vpt2 Rec fill
       2 copy vpt Square fill
       Bsquare } bind def
/S8 { BL [] 0 setdash 2 copy vpt sub vpt Square fill Bsquare } bind def
/S9 { BL [] 0 setdash 2 copy vpt sub vpt vpt2 Rec fill Bsquare } bind def
/S10 { BL [] 0 setdash 2 copy vpt sub vpt Square fill 2 copy exch vpt sub exch vpt Square fill
       Bsquare } bind def
/S11 { BL [] 0 setdash 2 copy vpt sub vpt Square fill 2 copy exch vpt sub exch vpt2 vpt Rec fill
       Bsquare } bind def
/S12 { BL [] 0 setdash 2 copy exch vpt sub exch vpt sub vpt2 vpt Rec fill Bsquare } bind def
/S13 { BL [] 0 setdash 2 copy exch vpt sub exch vpt sub vpt2 vpt Rec fill
       2 copy vpt Square fill Bsquare } bind def
/S14 { BL [] 0 setdash 2 copy exch vpt sub exch vpt sub vpt2 vpt Rec fill
       2 copy exch vpt sub exch vpt Square fill Bsquare } bind def
/S15 { BL [] 0 setdash 2 copy Bsquare fill Bsquare } bind def
/D0 { gsave translate 45 rotate 0 0 S0 stroke grestore } bind def
/D1 { gsave translate 45 rotate 0 0 S1 stroke grestore } bind def
/D2 { gsave translate 45 rotate 0 0 S2 stroke grestore } bind def
/D3 { gsave translate 45 rotate 0 0 S3 stroke grestore } bind def
/D4 { gsave translate 45 rotate 0 0 S4 stroke grestore } bind def
/D5 { gsave translate 45 rotate 0 0 S5 stroke grestore } bind def
/D6 { gsave translate 45 rotate 0 0 S6 stroke grestore } bind def
/D7 { gsave translate 45 rotate 0 0 S7 stroke grestore } bind def
/D8 { gsave translate 45 rotate 0 0 S8 stroke grestore } bind def
/D9 { gsave translate 45 rotate 0 0 S9 stroke grestore } bind def
/D10 { gsave translate 45 rotate 0 0 S10 stroke grestore } bind def
/D11 { gsave translate 45 rotate 0 0 S11 stroke grestore } bind def
/D12 { gsave translate 45 rotate 0 0 S12 stroke grestore } bind def
/D13 { gsave translate 45 rotate 0 0 S13 stroke grestore } bind def
/D14 { gsave translate 45 rotate 0 0 S14 stroke grestore } bind def
/D15 { gsave translate 45 rotate 0 0 S15 stroke grestore } bind def
/DiaE { stroke [] 0 setdash vpt add M
  hpt neg vpt neg V hpt vpt neg V
  hpt vpt V hpt neg vpt V closepath stroke } def
/BoxE { stroke [] 0 setdash exch hpt sub exch vpt add M
  0 vpt2 neg V hpt2 0 V 0 vpt2 V
  hpt2 neg 0 V closepath stroke } def
/TriUE { stroke [] 0 setdash vpt 1.12 mul add M
  hpt neg vpt -1.62 mul V
  hpt 2 mul 0 V
  hpt neg vpt 1.62 mul V closepath stroke } def
/TriDE { stroke [] 0 setdash vpt 1.12 mul sub M
  hpt neg vpt 1.62 mul V
  hpt 2 mul 0 V
  hpt neg vpt -1.62 mul V closepath stroke } def
/PentE { stroke [] 0 setdash gsave
  translate 0 hpt M 4 {72 rotate 0 hpt L} repeat
  closepath stroke grestore } def
/CircE { stroke [] 0 setdash 
  hpt 0 360 arc stroke } def
/Opaque { gsave closepath 1 setgray fill grestore 0 setgray closepath } def
/DiaW { stroke [] 0 setdash vpt add M
  hpt neg vpt neg V hpt vpt neg V
  hpt vpt V hpt neg vpt V Opaque stroke } def
/BoxW { stroke [] 0 setdash exch hpt sub exch vpt add M
  0 vpt2 neg V hpt2 0 V 0 vpt2 V
  hpt2 neg 0 V Opaque stroke } def
/TriUW { stroke [] 0 setdash vpt 1.12 mul add M
  hpt neg vpt -1.62 mul V
  hpt 2 mul 0 V
  hpt neg vpt 1.62 mul V Opaque stroke } def
/TriDW { stroke [] 0 setdash vpt 1.12 mul sub M
  hpt neg vpt 1.62 mul V
  hpt 2 mul 0 V
  hpt neg vpt -1.62 mul V Opaque stroke } def
/PentW { stroke [] 0 setdash gsave
  translate 0 hpt M 4 {72 rotate 0 hpt L} repeat
  Opaque stroke grestore } def
/CircW { stroke [] 0 setdash 
  hpt 0 360 arc Opaque stroke } def
/BoxFill { gsave Rec 1 setgray fill grestore } def
end
}}%
\begin{picture}(2160,1080)(0,0)%
{\GNUPLOTspecial{"
gnudict begin
gsave
0 0 translate
0.100 0.100 scale
0 setgray
newpath
1.000 UL
LTb
400 300 M
63 0 V
1647 0 R
-63 0 V
400 436 M
63 0 V
1647 0 R
-63 0 V
400 572 M
63 0 V
1647 0 R
-63 0 V
400 708 M
63 0 V
1647 0 R
-63 0 V
400 844 M
63 0 V
1647 0 R
-63 0 V
400 980 M
63 0 V
1647 0 R
-63 0 V
400 300 M
0 63 V
0 617 R
0 -63 V
742 300 M
0 63 V
0 617 R
0 -63 V
1084 300 M
0 63 V
0 617 R
0 -63 V
1426 300 M
0 63 V
0 617 R
0 -63 V
1768 300 M
0 63 V
0 617 R
0 -63 V
2110 300 M
0 63 V
0 617 R
0 -63 V
1.000 UL
LTb
400 300 M
1710 0 V
0 680 V
400 980 L
0 -680 V
1.000 UL
LT0
400 980 M
17 0 V
18 -1 V
17 -1 V
17 -1 V
17 -2 V
18 -2 V
17 -2 V
17 -3 V
17 -3 V
18 -4 V
17 -3 V
17 -5 V
18 -4 V
17 -5 V
17 -5 V
17 -5 V
18 -5 V
17 -6 V
17 -6 V
17 -6 V
18 -6 V
17 -7 V
17 -6 V
18 -7 V
17 -7 V
17 -7 V
17 -7 V
18 -8 V
17 -7 V
17 -7 V
17 -8 V
18 -7 V
17 -8 V
17 -8 V
18 -7 V
17 -8 V
17 -8 V
17 -7 V
18 -8 V
17 -8 V
17 -8 V
17 -7 V
18 -8 V
17 -7 V
17 -8 V
18 -8 V
17 -7 V
17 -8 V
17 -7 V
18 -7 V
17 -8 V
17 -7 V
17 -7 V
18 -7 V
17 -7 V
17 -7 V
18 -7 V
17 -7 V
17 -7 V
17 -7 V
18 -7 V
17 -6 V
17 -7 V
17 -6 V
18 -7 V
17 -6 V
17 -7 V
18 -6 V
17 -6 V
17 -6 V
17 -6 V
18 -6 V
17 -6 V
17 -6 V
17 -6 V
18 -5 V
17 -6 V
17 -6 V
18 -5 V
17 -5 V
17 -6 V
17 -5 V
18 -5 V
17 -6 V
17 -5 V
17 -5 V
18 -5 V
17 -5 V
17 -5 V
18 -5 V
17 -5 V
17 -4 V
17 -5 V
18 -5 V
17 -4 V
17 -5 V
17 -4 V
18 -5 V
17 -4 V
1.000 UL
LT1
417 980 M
18 -1 V
17 0 V
17 -1 V
17 -2 V
18 -1 V
17 -2 V
17 -3 V
17 -2 V
18 -3 V
17 -3 V
17 -3 V
18 -4 V
17 -3 V
17 -4 V
17 -4 V
18 -4 V
17 -5 V
17 -4 V
17 -5 V
18 -5 V
17 -5 V
17 -5 V
18 -5 V
17 -5 V
17 -6 V
17 -5 V
18 -5 V
17 -6 V
17 -5 V
17 -6 V
18 -6 V
17 -5 V
17 -6 V
18 -5 V
17 -6 V
17 -5 V
17 -6 V
18 -5 V
17 -6 V
17 -5 V
17 -5 V
18 -6 V
17 -5 V
17 -5 V
18 -5 V
17 -6 V
17 -5 V
17 -5 V
18 -5 V
17 -5 V
17 -4 V
17 -5 V
18 -5 V
17 -5 V
17 -4 V
18 -5 V
17 -4 V
17 -5 V
17 -4 V
18 -4 V
17 -5 V
17 -4 V
17 -4 V
18 -4 V
17 -4 V
17 -4 V
18 -4 V
17 -4 V
17 -3 V
17 -4 V
18 -4 V
17 -3 V
17 -4 V
17 -3 V
18 -4 V
17 -3 V
17 -3 V
18 -4 V
17 -3 V
17 -3 V
17 -3 V
18 -3 V
17 -3 V
17 -3 V
17 -3 V
18 -3 V
17 -3 V
17 -3 V
18 -3 V
17 -2 V
17 -3 V
17 -3 V
18 -2 V
17 -3 V
17 -2 V
17 -3 V
18 -2 V
17 -3 V
stroke
grestore
end
showpage
}}%
\put(1255,50){\makebox(0,0){$k$ [fm$^ {-1}$]}}%
\put(100,640){%
\special{ps: gsave currentpoint currentpoint translate
270 rotate neg exch neg exch translate}%
\makebox(0,0)[b]{\shortstack{R}}%
\special{ps: currentpoint grestore moveto}%
}%
\put(2110,200){\makebox(0,0){10}}%
\put(1768,200){\makebox(0,0){8}}%
\put(1426,200){\makebox(0,0){6}}%
\put(1084,200){\makebox(0,0){4}}%
\put(742,200){\makebox(0,0){2}}%
\put(400,200){\makebox(0,0){0}}%
\put(350,980){\makebox(0,0)[r]{1}}%
\put(350,844){\makebox(0,0)[r]{0.9}}%
\put(350,708){\makebox(0,0)[r]{0.8}}%
\put(350,572){\makebox(0,0)[r]{0.7}}%
\put(350,436){\makebox(0,0)[r]{0.6}}%
\put(350,300){\makebox(0,0)[r]{0.5}}%
\end{picture}%
\endgroup
 